\documentstyle[12pt]{article}

\begin{document}

\begin{center}
\Large{{\bf Time-dependent oscillator with Kronig-Penney excitation}}
\end{center}
\begin{center}
V.V. Dodonov, O. V. Man'ko, V. I. Man'ko\\
Lebedev Physical Institute, Leninsky Prospect 53, 117924 Moscow, Russian Federation\\
\end{center}

\begin{abstract}
Exact solutions of the time-dependent Schr\"odinger equation for a quantum oscillator subject to periodical frequency $\delta$-kicks are obtained. We show that the oscillator occurs in the squeezed state and calculate the corresponding squeezing coefficients and the energy increase rate in terms of Chebyshev polynomials.
\end{abstract}

The problem of quantum oscillator with a time-dependent frequency
was solved in refs.~\cite{1,2,3}. It was shown that the wave function
and, consequently, all physical characteristics of the oscillator
can be expressed in terms of the solution of the classical equation
of motion
\begin{equation}\label{1}
\dot\epsilon(t)+\omega^2(t)\epsilon(t)=0.
\end{equation}
(for detailed review see ref.\cite{disert4a}). The only remaining problem is to find explicit expression for the $\epsilon$-function. A lot of papers were devoted to that in the last two decades. One can find an extensive list of references, e.g., in ref. \cite{disert4a}. Here we consider the interesting case of periodically kicked oscillator, where the frequency depends on time as follows
%Let us consider, following~\cite{DMMKronig-Penney}, an interesting
%example of oscillator subject to periodical frequency delta-kicks
%(Kronig--Penney parametric excitation). The frequency depends on
%time as follows,
\begin{equation}\label{disert4.1}
\omega^2(t)=\omega^2_0-2\kappa\sum_{k=0}^{n-1}\delta(t-k t),
\end{equation}
where $\omega_0$ is constant frequency, $\delta$ - is the 
Dirac delta-function. So we have the equation for $\epsilon(t)$-function 
\begin{equation}\label{disert4.1a}
-\frac{\ddot\epsilon(t)}{2}+\kappa\sum_{k=0}^{N-1}\delta(t-k\tau)\epsilon(t)=\frac{1}{2}\omega_0^2\epsilon(t).
\end{equation}
Obviously, upon the substitutions $t\rightarrow
x,\,\epsilon\rightarrow \psi,\,\omega_0^2\rightarrow E$, this
equation~(\ref{disert4.1a}) coincides with the equation for the wave
function of a quantum particle of unity mass in a Kronig-Penney
potential (a sequence of delta-potentials)~\cite{disert180}-\cite{disert98}. The case $\kappa>0$ corresponds to
sequence of $\delta$-barrier, while the  case $\kappa<0$ relates to
sequence of $\delta$-wells. %The solution of this equation is well
%known and is considered in a lot of papers (see for example,
%paper~\cite{disert95}-\cite{disert99}). 
For every time interval $(k-1)\tau<t<k \tau$ the solution for~(\ref{1},\ref{disert4.1}) is given by
$$\epsilon_k(t)=A_k e^{i\omega_0 t}+B_k e^{-i\omega_0 t}, \quad k=0,..., N. $$
Due to continuity conditions
\begin{equation}\label{PPUN}
\epsilon_{k-1}(k\tau)=\epsilon_k(k\tau),\quad
-\frac{1}{2}\left[\dot\epsilon_k(k\tau)
-\dot\epsilon_{k-1}(k\tau)\right]+\kappa\epsilon_{k-1}(k\tau)=0.
\end{equation}
(the latter is obtained by integrating eq.~(\ref{1}) over the
infinitely small interval $k\tau-0<t<k\tau+0$, the coefficients
$A_k$ and $B_k$ must satisfy some relations which can be represented
in matrix form
\[
\pmatrix{A_k e^{i\omega_0 t_k}\cr B_k e^{-i\omega_0 t_k}}=\pmatrix{1-\frac{i\kappa}{\omega_0}&-\frac{i\kappa}{\omega_0}\cr
\frac{i\kappa}{\omega_0}&1+\frac{i\kappa}{\omega_0}}\pmatrix{A_{k-1} e^{i\omega_0 t_k}\cr B_{k-1} e^{-i\omega_0 t_k}}.
\]
After a sequence of $\delta$-kicks the coefficients
$A_n,\,B_n$ are connected with the initial ones $A_0,\,B_0$ through
the equation
\begin{equation}\label{MT1}
\pmatrix{A_n\cr B_n}=S^{(n)}\pmatrix{A_0\cr B_0}=T^{-(n-1)}(M T
)^n\pmatrix{A_0\cr B_0},
\end{equation}
with the matrices $T,\,M$ given by
\[
M=\pmatrix{1-\frac{i\kappa}{\omega_0}&-\frac{i\kappa}{\omega_0}\cr \frac{i\kappa}{\omega_0}&1+\frac{i\kappa}{\omega_0}},
\quad T=\pmatrix{e^{i\omega_0 \tau}&0\cr 0&e^{-i\omega_0 \tau}}.
\]
The matrix
\[
S^{(1)}\equiv S=M T=\pmatrix{\Big(1-\frac{i\kappa}{\omega_0}\Big)e^{i\omega_0\tau}&
-\frac{i\kappa}{\omega_0}e^{-i\omega_0\tau} \cr \frac{i\kappa}{\omega_0}e^{i\omega_0\tau}&
\Big(1+\frac{i\kappa}{\omega_0}\Big)e^{-i\omega_0\tau}}
\]
is unimodular: $\mbox{det}S=1$. It can be shown (see,
e.g., ref.~\cite{disert181}), that all the powers of two-dimensional
unimodular matrices can be expressed as linear combinations of the
matrices can be expressed as linear combinations of the matrices
themselves and the identity matrix, the coefficients being Chebyshev
polynomials of the second kind whose arguments are expressed in
terms of the traces of the initial matrices ($E$ is the identity
matrix)
\begin{equation}\label{chebyshev}
S^{n}=U_{n-1}(\cos\phi)S-U_{n-2}(\cos\phi)E,\quad \cos\phi=\frac{1}{2}\mbox{Tr}S=\frac{\chi}{2}.
\end{equation}
We use the following definitions of the Chebyshev
polynomials~\cite{disert182}
\begin{equation}\label{disert4.1c}
U_n(\cos\phi)=\frac{\sin\left[\left(n+1\right)\phi\right]}{\sin\phi},\quad
U_n(\chi)=\frac{\left[\chi+\left(\chi^2-4\right)^{\frac{1}{2}}\right]^{n+1}-
\left[\chi-\left(\chi^2-4\right)^{\frac{1}{2}}\right]^{n+1}}{2^{n+1}\left(\chi^2-4\right)^{\frac{1}{2}}}.
\end{equation}
In the case under study
\[
\frac{\chi}{2}=\cos\phi=\cos\omega_0\tau+\frac{\kappa}{\omega_0}\sin\omega_0\tau.
\]
Thus the matrix elements of the matrix $S^{(n)}$ can be written as
follows
\[
S_{11}^{(n)}=\Big(1-\frac{i\kappa}{\omega_0}\Big)U_{n-1}\big(\frac{\chi}{2}\big)e^{-i\omega_0(n-2)\tau}-
U_{n-2}\big(\frac{\chi}{2}\big)e^{-i\omega_0(n-1)\tau},
\]
\[
S_{12}^{(n)}=-\frac{i\kappa}{\omega_0}U_{n-1}\big(\frac{\chi}{2}\big)e^{-i\omega_0 n \tau},
\quad S_{21}^{(n)}=\frac{i\kappa}{\omega_0}U_{n-1}\big(\frac{\chi}{2}\big)e^{i\omega_0 n\tau},
\]
\[
S_{22}^{(n)}=\Big(1+\frac{i\kappa}{\omega_0}\Big)U_{n-1}\big(\frac{\chi}{2}\big)e^{i\omega_0(n-2)\tau}-
U_{n-2}\big(\frac{\chi}{2}\big)e^{i\omega_0(n-1)\tau}.
\]
If
%\begin{equation}\label{initialcond}
$\epsilon(-0)=1,\quad \dot\epsilon(-0)=i\omega_0 $
%\end{equation}
at the initial time, then $A_0=1,\,B_0=0,$ so $A_n=S_{11}^{(n)},
B_n=S_{21}^{(n)}.$
If at the initial time, $t<0$, a quantum oscillator was in a
coherent state, a parametric excitation will transform it into a
squeezed correlated state with the coordinate variance 
%\[
$\sigma_{xx}(t)=\frac{\hbar|\epsilon|^2}{2m\omega_0}$ \cite{disert4a}. 
%\]
So after a sequence of $\delta$-kicks one has
\begin{eqnarray*}
&&\sigma_{xx}(t)=\frac{\hbar}{2m\omega_0}\Big(U_{n-1}^2+U_{n-2}^2+
\frac{2\kappa}{\omega_0}U_{n-1}^2\sin[2\omega_0(t-(n-1)\tau)]-\chi U_{n-1}U_{n-2}\Big.\nonumber\\
&&+\Big.\frac{4\kappa^2}{\omega_0^2}U_{n-1}^2\sin^2[\omega_0(t-(n-1)\tau)]
-\frac{2\kappa}{\omega_0}U_{n-1}U_{n-2}\sin[2\omega_0(t-(n-\frac{1}{2})\tau)]\Big).\nonumber
\end{eqnarray*}
The correlation between coordinate and momentum in a correlated
squeezed state is not equal to zero \cite{disert4a} 
\[
|\sigma_{x p}(t)=\frac{1}{2}\hbar\left(1-|\epsilon\dot\epsilon|^2/\omega^2\right)^{1/2}.
\]
After a sequence of $\delta$-kicks it has the form 
\begin{eqnarray*}
&&\sigma_{x p}=\frac{\hbar}{2}\Big\{\Big[\big(1 +2\frac{\kappa^2}{\omega_0^2}\big)U_{n-1}^2+U_{n-2}^2-
\chi U_{n-1}U_{n-2}\Big]^2-\Big[\frac{2\kappa}{\omega_0}U_{n-1}^2\sin[2\omega_0(t-(n-1)\tau)]\Big.\Big. \nonumber\\
&&\Big.\Big.-\frac{2\kappa^2}{\omega_0^2}\cos[2\omega_0(t-(n-1)\tau)]-
\frac{2\kappa}{\omega_0}U_{n-1}U_{n-2}\sin[2\omega_0(t-(n-\frac{1}{2})\tau)]\Big]^2-1\Big\}^{\frac{1}{2}}.\nonumber
\end{eqnarray*}
The dimensionless energy of the quantum fluctuations $W=2E/\hbar\omega_0$ (normalized by
the ground state energy $\hbar\omega_0/2$) after the $n$ kicks equals
\begin{equation}\label{disert4.2}
W=(m \omega_0/\hbar)\sigma_{x x}(t)+\sigma_{p p}(t)/\hbar m\omega_0=
U_{n-1}^2\big(\frac{\chi}{2}\big)+U_{n-2}^2\big(\frac{\chi}{2}\big)
+\frac{4\kappa^2}{\omega_0^2}U_{n-1}^2\big(\frac{\chi}{2}\big)
-\chi U_{n-1}\big(\frac{\chi}{2}\big)U_{n-2}\big(\frac{\chi}{2}\big).
\end{equation}
As was shown~\cite{disert57}, the maximum coefficients of the
squeezing and of the correlation are expressed through the
dimensionless energy of the fluctuations as follows,
\[
r_{max}=(1-1/W^2)^{1/2}, \quad k_{max}^2=\frac{1+(1-1/W^2)^{1/2}}{1-(1-1/W^2)^{1/2}}.
%\frac{\sqrt{4E^2-1}}{2E},\quad k_{max}^2=\frac{2E+\sqrt{4E^2-1}}{2E-\sqrt{4E^2-1}}.
\]
\noindent In the case when the parameter $\chi/2$ does not belong to
the interval $[-1,1]$, the asymptotic formula for the Chebyshev
polynomials~\cite{disert182} with $n>>1$,
\[
U_n(x)=\frac{\sqrt{n}}{2\sqrt{n-1}}\big(x^2-1\big)^{-\frac{1}{2}}
\Big[\sqrt{x+1}+\sqrt{x-1}\Big]\Big[x+\sqrt{x^2-1}\Big]^{n-\frac{1}{2}},
\]
gives rise to the asymptotic expression for the dimensionless energy of the fluctuations (we suppose  $\chi>0$),
%\noindent where $x=\chi/2$. If also $(n\rightarrow\infty)$, then one
%has for~(\ref{disert4.2}) asymptotic formulas
\[
W=\frac{4\kappa^2\big[x+\big(x^2-1\big)^{\frac{1}{2}}\big]^{2n}}{\omega_0^2\big(x^2-1\big)},\quad x=\chi/2.
\]
If we choose the simplest case, when the interval between the kicks satisfies the condition 
$\omega_0\tau=\frac{\pi}{2}+2\pi k,\quad k=0,1,\,...\,$, the arguments of the Chebyshev polynomials are equal to
$\kappa/\omega_0$. When the strength of the $delta$-kicks $\kappa$
is larger than the constant frequency $\omega_0$, one gets the
asymptotic formula
\[
W=\frac{4\kappa^2\big[\frac{\kappa}{\omega_0}+\big(\frac{\kappa^2}{\omega_0^2}-1\big)^{\frac{1}{2}}\big]^{2n}}
{\omega_0^2\big(\frac{\kappa^2}{\omega_0^2}-1\big)}.
\]
In the case when $\kappa/\omega_0>>1$, the energy of the
fluctuations increases exponentially with the number of kicks
\[
W=4\big(2\kappa/\omega_0\big)^{2n},\quad n>>1.
\]
If the $\delta$-kicks take place at moments given by
$\omega_0\tau=2\pi m,\,m=0,1,\,...\,$, then the argument of the
Chebyshev polynomials is equal to unity independently of the
strength of the $\delta$-kicks. Then the energy increases much more
slowly,
\[
W=\big(1+\frac{4\kappa^2n^2}{\omega_0^2}\big).
\]
In the case when the strength of the $\delta$-kicks is
small, $\kappa/\omega_0<<1$, and the interval between the kicks is
given by 
%\[
$\omega_0\tau=\frac{\pi}{2}+2\pi k,\quad k=0,1,\,...\,,$ 
%\] 
the following inequality for the maximum total energy can
be obtained,
\[
W_{max}<\Big(1+\frac{4n^2\kappa^2}{\omega_0^2}\Big).
\]
%\noindent Let us consider that the oscillator is on $N$--th
%stationary level at $t<0$, then for time $t>(n-1)\tau$ the
%transition probability on $M$--th stationary level due to parametric
%excitation is of the form~\cite{disert4a}
%\begin{equation}\label{24a}
%W_N^M(t)=\frac{N!}{M!\sqrt{E(t)+\frac{1}{2}}}\Big(P_{\frac{M+N}{2}}^{\frac{M-N}{2}}
%\Big[\big(E(t)+\frac{1}{2}\big)^{-\frac{1}{2}}\Big]\Big)^2 ,\quad N\leq M,
%\end{equation}
%\noindent in the case of $n$ $\delta$-kicks of frequency is
%\[
%\begin{equation}\label{24}
%W_N^M=\frac{N!\sqrt2 }{M!\sqrt{\big(1+\frac{4\kappa^2}{\omega_0^2}\big)U_{n-1}^2
%+U_{n-2}^2-\chi U_{n-1}U_{n-2}+1}}
%\Big[ P_{\frac{M+N}{2}}^{\frac{M-N}{2}}\Big(\frac{\sqrt2}{\sqrt{\big(1+\frac{4\kappa^2}%{\omega_0^2}\big)U_{n-1}^2
%+U_{n-2}^2-\chi U_{n-1}U_{n-2}+1}}\Big)\Big]^2,
%\end{equation}
%\]
%\noindent where $P_{\alpha}^{\beta}$ - Legender polynomials.

\end{document}